\documentclass[twocolumn,showpacs,preprintnumbers,amsmath,amssymb]{revtex4}
\usepackage{epsfig}
\usepackage{color}
\usepackage{graphicx}
\begin{document}

\def \be{\begin{equation}}
\def \ba{\begin{eqnarray}}
\def \ea{\end{eqnarray}}
\def \ee{\end{equation}}
\
\title{Random sequential adsorption of shrinking or spreading particles }
\author{Arsen V. Subashiev, Serge Luryi}
\affiliation{Department of Electrical and Computer Engineering,
State University of New York at Stony Brook, Stony Brook, NY,
11794-2350}
%

\begin{abstract} We present a model of one-dimensional irreversible
adsorption in which particles once adsorbed immediately shrink to a
smaller size or expand to a larger size. Exact solutions for the
fill factor and the particle number variance as a function of the size
change are obtained. Results are compared with approximate
analytical solutions.
\end{abstract}
\maketitle \nopagebreak

\section{Introduction}
Random sequential adsorption (RSA) is an attractive model for a
number of physical phenomena, including such different application
as information processing \cite{coff} and particle branching in
impact ionization \cite{ino}. The simplest example of RSA is the car
parking problem (CPP). Of interest is the average number of
particles ("cars") adsorbed on a long line, as well as the variance
of this number (see Refs. \cite{evans,Talbot} for the review). In
what follows, we shall use the term \emph{standard RSA} for the
classical 1-dimensional model corresponding to particles of fixed
size that arrive randomly on a line and are deposited if empty space
is available and rejected otherwise. Extensions of this model
include RSA with particles expanding in the adsorption process
\cite{viot}, two-size particle adsorption \cite{hassan,Araujo}, and
also RSA with an arbitrary particle-size distribution function
\cite{mao}.

Few of these models, especially physically relevant ones, have an
exact solution. As a rule, only the fill factor is determined.
However, for a number of applications fluctuations are of major
importance. An example of such an application is the very
important practical problem of particle energy branching (PEB)
where high-energy particle propagates in an absorbing medium and
multiplies producing secondary electron-hole (e-h) pairs. The
energy distribution of secondary particles is random to a good
approximation. Multiplication proceeds so long as the particle
energy is above the impact ionization threshold \cite{Spieler}.
This connection was noticed as early as in 1965 by W. van
Roosbroek \cite{rusb}. The extension of RSA model proposed in
\cite{rusb}, known as the "crazy carpenter model" was further
exploited in \cite{ino,Alig}.

The PEB process can be considered in terms of a CPP if one
identifies the initial particle kinetic energy with an available
parking length and the pair creation energy as the car size. Full
equivalence to CPP requires further that only one of the secondary
particles takes on significant energy, otherwise one has to
consider simultaneous random parking of two cars in one event
\cite{gamma}.

The number of created electron-hole pairs $N$ in PEB serves to
evaluate the initial energy. Variance of this number limits the
accuracy of energy measurements. Both the yield $\overline{N}$ and
the e-h pair variance ${\rm var} (N)=\overline{(N-\overline{N})^2}$
are proportional to initial energy. The ratio of the e-h pair
variance to the yield is called the Fano factor \cite{Fano}. The
Fano factor $\Phi$ (for a Poisson-distributed $N$ one has $\Phi =
1$) is a parameter that quantifies the energy resolution of
high-energy particle detectors. For semiconductor crystals, the PEB
problem has many additional complications due to phonon losses, as
well as features in energy dependencies of the particle density of
states and the impact ionization matrix element.

Earlier attempts to evaluate the Fano factor for the PEB problem in
semiconductors employed widely different approaches (compare
\cite{rusb,Spieler} and \cite{Klein}). To obtain agreement with the
experimentally observed $\Phi \approx 0.1$ (for semiconductors),
different fairly rude and unjustified assumptions were made, so that
the numerical coincidence is of little value. Also available are
numerical calculations \cite{Alig,casanova}. However, the relative
importance of various factors (e.g. phonon contribution to $\Phi$)
remains questionable within the numerical models, while the
precision of results is difficult to assess. Evaluation of the Fano
factor is important for predicting the energy resolution of
detectors, especially those based on new materials and new
principles \cite{Serge}.

An important aspect of the impact ionization process is the fact
that the threshold impact ionization energy is usually larger than
the minimum energy $E_g$ needed for e-h pair creation. This
difference arises from kinematic restrictions imposed by momentum
conservation, so that, e.g., for equal effective masses of electrons
and holes the minimal energy of pair production is $3/2 E_g$. In
crystals with non-equal electron and hole masses the threshold can
vary from $E_g$ to $2E_g$. In terms of the RSA problem this is
equivalent to the particle shrinking immediately after the
adsorption (parking), making larger available length for the next
adsorption event.

For the PEB problem only particle shrinking is pertinent. However,
for a general RSA problem  both the shrinking and the expansion of
particles are relevant, due to such factors as, e.g., repulsion
between particles and surface attraction.

In this paper, we use a recursive approach to consider the general
RSA problem for particles that either shrink or expand immediately
upon adsorption. We present an analytical solution for both the
average filling factor and its variance. The results are compared
with approximate solutions obtained by the methods employed in
earlier approaches to the energy branching problem
\cite{ino,rusb,Klein,Alig}  and, for expanding particles, also
with the results of a kinetic approach used in Ref. \cite{viot}.
Exact solutions presented here allow to assess the errors
introduced by the adopted approximations in the evaluation of both
the yield and the variance.

\section{Statistics of adsorption}

In this section we develop an analytical model of adsorption for
shrinking or expanding particles. We shall be using the recursive
technique \cite{mao}. Besides being the most direct approach, this
technique has the further advantage of being applicable for the
evaluation of both the average filling factor and the variance. The
recursive approach has been confirmed by direct Monte-Carlo computer
simulations\cite{mao,rusb,casanova} for similar problems.

Consider adsorption  of a particle on an initially empty line of
progressively growing length $x$. The final (after shrinking or
expansion) size of the adsorbed particle is taken equal unity.
Denote by $w_x$ the random variable corresponding to the total
wasted length for some configuration of adsorbed particles in the
\emph{jamming limit}, when all gaps are smaller than the minimal
length needed for a particle to be adsorbed. In the process of
sequential adsorption deposition of additional particle generates
new gaps with the same distribution of the gap size. Equations for
the expected values of the moments of waste distribution can be
obtained using moment generation function, $\Psi(\lambda,w_x)={\bf
E}\exp(\lambda w_x)$ \cite{MGF}. Moments of $w_x$ are obtained by
differentiating $\Psi$ with respect to $\lambda$. When a particle
is adsorbed into gap $x$, two new gaps appear, the size of the
gaps being $y$ and $x-1-y$. Therefore for $\Psi$ we have \be
\Psi(\lambda,w_x)=<\Psi(\lambda,w_y)\Psi(\lambda,w_{x-1-y})> .
\label{MGF1} \ee In this equation, the angular brackets denote
averaging over the distribution of $y$, characterized by a
one-particle gap distribution function (OPDF),  $\rho(y|x)$, that
equals to the probability density of creating an interval $y$ in a
one-particle deposition into the line $x$, viz. \be
<\Psi(\lambda,w_y)>=\int_{y_{min}}^{y_{max}}\Psi(\lambda,w_y)
\rho(y|x) dy. \label{OPD} \ee The randomness in $y$ is induced by
a two-step random process, (i) the random choice of the adsorption
coordinate and (ii) the subsequent size change, which depends on
the adopted model of shrinkage/expansion.

Using Eq. (\ref{MGF1}) in calculating the first derivative of
$\Psi(\lambda,w_x)$ at $\lambda=0$, we find the first moment (the mean value) of
waste, ${\bf E} w_x \equiv w(x)$, in the form
\be
 w(x)=< w(y)>+< w(x-1-y)>= 2< w(y)>,
\label{MastW1}
\ee the latter equation being due to symmetry. The
second moment, $u(x)\equiv {\bf E} w_x^2,$ needed for evaluation of
variance, is obtained by taking the second derivative, \be u(x) =
2<u(y)>+2<w(y)w(x-1-y)>\label{MastU1}. \ee Equivalent equations can
be derived for the mean filled length $f(x)$, which for a unit
particle size equals the number of particles adsorbed on the line of
length $x$. The random variables $w_x$ and $f_x$ are complementary
in the sense that $w_x=x-f_x$ and, therefore, equations for $f(x)$
can be obtained from Eqs. (\ref{MastW1}) and (\ref{MastU1}) by the
substitution $w(x)=x-f(x)$.

\begin{figure} 
\epsfig{figure=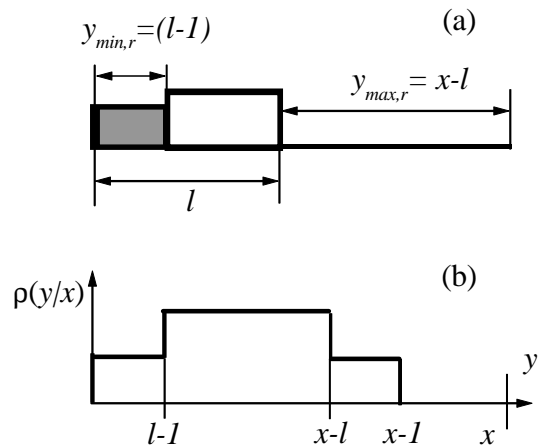,width=7.3cm,height=6.3cm}
\caption{Illustration (a) of particle adsorption near the edge of
a gap of size $x$ in the model where the particle is assumed to
shrink upon adsorption (from size $l$ to unit size) by randomly
retracting one of its ends (the retracted part is marked by the
dark shading, the final state is shown as a white brick of unit
length); $y_{min,r}$ and $y_{max,r}$ are minimal and maximal gaps
created by particles adsorbed at left end of the length $x$ and
retracting to right endpoint ; part (b) shows the one particle gap
probability distribution function or OPDF, $\rho(y|x)$. }
\label{1f}
\end{figure}
For the standard RSA problem the OPDF is a homogeneous distribution
of $y$ in the interval $\{0,x-1\}$). The corresponding equations
were obtained in \cite{coff,rusb} in the context of RSA  and in
\cite{rusb,ino} for the energy branching problem.

An alternative approach to RSA  is to consider a kinetic (or rate)
equation that describes the sequential deposition of particles. In
the kinetic approach the gap size distribution function $G(x,t)$
representing the average density of voids of the length between
$x$ and $x+dx$ at a time $t$ obeys the equation \cite{Hassan2} \ba
\frac {\partial G(x,t)}{\partial t} = - G(x,t)\int_0^x ds p(s) \int_0^{x-s}dy F(y,x-y-s|s) \nonumber \\
+2  \int_x^{\infty}dy G(y,t) \int_0^{y-x} ds p(s) F(x,y-x-s|s).
\hspace{0.5cm} \label{1Hass} \ea Here $p(s)$ is the adsorbed
particle distribution function, and $F(y,x-y-s|s)$ is the
deposition probability, that determines the average rate, \be
R(y|x) = \int_0^x ds p(s)  F(y,x-y-s|s), \ee at which the initial
length $x$ is destroyed by the deposition of a particle producing
a gap $y$. One can see that for a fixed $x=x_0$  the rate $R$ and
our OPDF are proportional to each other, $\rho(y|x_0) \propto
R(y|x_0)$, the only difference being due to the fact that
$\rho(y|x_0)$ is normalized to unity.

For the case when the randomly adsorbed particles change their size,
the OPDF depends on the particular model of size transformation in the
adsorbed state. Several such models are discussed below.

{\bf Shrinking particles.} First, we consider a model in which
particles of initial length $l>1$ shrink to the length of unity by
randomly retracting one of their endpoints (either on the left or on
the right with equal probability).

In this model, no particles are adsorbed for $x<l$, i.e., small
intervals are wasted entirely, $w(x)=x$. When the interval length
$x$ reaches $l$, then two gaps of size $(l-1)/2$ are created. For an
arbitrary $x>l,$ adsorption of a particle creates two new gaps $y$
and $x-1-y$. Note that the probability of creating gaps near the
edges of the initial interval (i.e., for $0<y<l-1$ and $x-l<y<x-1$)
is twice smaller, since they get contributions only from particles
that retract the right endpoint (at the right side of the interval
$x$) and the left endpoint (at the left side of $x$), respectively,
see Fig. 1. This model can be used for the PEB problem, to account
for the effect of lower density of states at low particle energies.

For sufficiently large intervals, $x>2l-1$, the mean waste is
described by the following equation
\ba w(x) =\frac{1}{x-l}
\left(\int_0^{l-1}w(y)dy +2\int_{l-1}^{x-l}w(y)dy \right.
\nonumber \\ \left. +\int_{x-l}^{x-1}w(y)dy\right)\hspace{2cm}
\label{MastWst3}
\ea
Note that Eq. (\ref{MastWst3}) does not work in the region $l
<x< 2l-1$, where the gap distribution $\rho(y|x)$ is simply uniform
within $y <x-l$ and $l-1 <y <x-1$. However, since the adsorption of the second
particle starts at $x>l+1$, knowledge of the variation of $w(x)$ at
small $x<l+1$ is sufficient to obtain an exact solution of Eq.
(\ref{MastWst3}) --- so long as $2l-1 < l+1$ (or $l<2$). In what
follows we shall confine ourselves to the case $l<2$.

According to (\ref{MastU1}) the equation for the second moment
$u(x)$ is obtained from Eq. (\ref{MastWst3}) by the substitution
$w(y) \rightarrow u(y)+w(y)w(x-1-y)$.

\begin{figure} 
\epsfig{figure=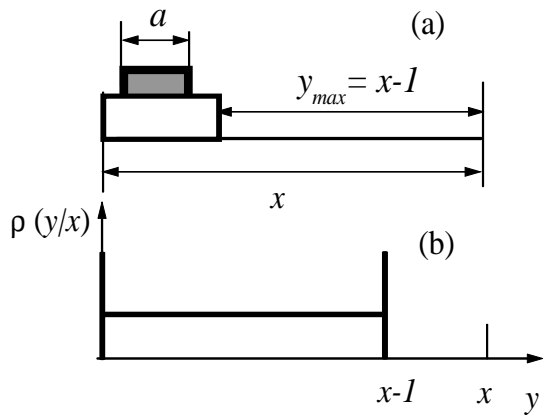,width=7.3cm,height=5.8cm} \caption
{Adsorption of expanding particles with the transformation rules
corresponding to a symmetric expansion $a \Rightarrow 1$ when size
permits and near an edge an asymmetric expansion to fill the
available space; (a) illustration of particle adsorption near the
edge of a gap of size $x$; (b) one-particle gap distribution
function, $\rho(y|x)$, shows $\delta$-function singularities at
$y=x-1$ and $y=0$.} \label{2f}
\end{figure}

Another possible shrinking model results when the adsorbed
particle is assumed to shrink symmetrically about its center.  In
this case the OPDF remains constant within the interval $(l-1)/2
\le y \le x-(l+1)/2$. This model accounts for the particle
repulsion in a model of hard disks, when the distance between the
particle centers can not be smaller than a certain length that
serves as their effective size. That length can often be larger
than the actual particle size used in the description of the
resulting coverage. A similar model reasonably describes the PEB
process in semiconductors, when both secondary particles produced
by impact ionization are of the same mass and acquire equal
kinetic energies.

With this OPDF, equation (\ref{MastW1}) takes the form \be w(x)
=\frac{2}{x-l}\int_{(l-1)/2}^{x-(l+1)/2}w(y)dy . \label{MastW2}\ee
Similarly, equation for the second moment $u(x)$ reads \ba u(x) =
\frac{2}{x-l}\int_{(l-1)/2}^{x-(l+1)/2}u(y)dy \hspace{3cm}
\nonumber
\\+\frac{2}{x-l}\int_{(l-1)/2}^{x-(l+1)/2}w(y)w(x-1-y)dy.\label{MastU2}
\ea

These equations are different from those obtained for the standard
RSA problem. The modification arises due to the edge effect and is
important at the jamming limit, when gaps are minimal. For particles
shrinking to the center, the edge effect results in stronger
restrictions on the gap size.

{\bf Expanding particles.} Here we consider a model in which
particles of initial length $a<1$ given sufficient space expand
symmetrically to unit length. If the particle is placed in a gap of
size $a<x<1$, it fills it completely. If it is placed in a gap $x>1$
near its edge, it expands asymmetrically to unit length. This model
represents size transformation due to an attractive force from the
surface for the case when interaction between the particles is
negligible.

To define the ODPF for this model, we note that given the initial
length $x$, coordinates of the centers of adsorbed particles are
homogeneously distributed within $a/2,x-a/2$. For particles adsorbed
not too close to the gap edges, the gap distribution function
remains uniform in the interval $0<y<x-1$. However, all adsorbed
particles whose centers fall within an interval ($a/2,1/2$) from an
edge will produce a gap of the same size $x-1$,  see Fig. 2, as well
as a "gap" of zero width. This results in a singularity at $y=x-1$
(and another one at $y=0$) in the gap distribution function
(existence of this singularity for expanding particles was already
noted in \cite{viot}). With this ODPF, Eq. (\ref{MastW1}) acquires
the form \be w(x) =\frac{2}{x-a}\int_0^{x-1}w(y)dy +\frac{1-a}{x-a}
w(x-1). \label{MastW3}\ee Similarly, the second moment
equation becomes \ba u(x) = \frac{2}{x-a}\int_0^{x-1}u(y)dy+\frac{1-a}{x-a}
u(x-1)\nonumber
\\+\frac{2}{x-a}\int_0^{x-1}w(y)w(x-1-y)dy. \label{MastU3}
\ea Note that the particle expansion and edge effect do not
influence the second term of Eq. (\ref{MastU3}), since the product
$w(y)w(x-1-y)=0$ vanishes near the edges. Nevertheless, we note that
both Eqs. (\ref{MastW3}) and (\ref{MastU3}) are modified by the edge
effect.

In all models of size transformation considered above the ODPF
remains constant within certain intervals. It is this feature that
enables exact solution of the recursion equations.

It is worthwhile to stress that due to the self-averaging nature
of the filled length (and the waste length) in the limit
$x\rightarrow \infty$ solution of the averaged (hence approximate)
recursion equations gives exact results. The same is true for the
kinetic approach in the limit $t \rightarrow \infty$. Both
approaches are, therefore, equivalent for the calculation of the
fill factor.

\section{filling factor }\label{fillFx}
We evaluate the filling factor in the so-called "jamming limit" ---
corresponding to the situation when every gap capable of adsorbing a
particle has done so. The average density of particles saturates in
the jamming limit.

{\bf Shrinking particles.} We shall first focus on the model of
particles shrinking by randomly retracting one of the endpoints. The
case of particles shrinking to the center will be discussed
subsequently. We have to solve Eq. (\ref{MastWst3}) with appropriate
boundary conditions that originate from the region of small
available adsorption interval $x$. Consider an initially empty line
of progressively growing length $x$. For $x < l$ the entire interval
is wasted, $w(x)=x$. For $l<x<l+1$ only one particle will be
adsorbed producing waste equal to $w(x)=x-1$. Clearly, for $x<l+1$
the waste is fixed and does not fluctuate. For $x>l+1$ the
probability that two particles will be sequentially adsorbed grows
steadily and so does the average waste, which also begins to
fluctuate. At large $x$ both the wasted length and the average
covered length grow linearly with $x$, so that the average waste per
particle or per unit adsorption interval remains constant.
\begin{figure} 
\epsfig{figure=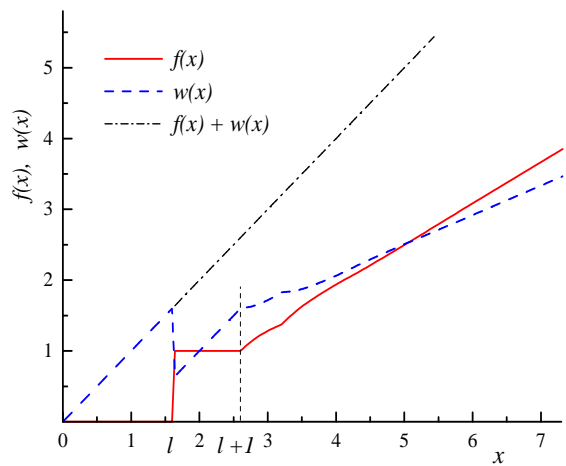,width=7.6cm,height=6.3cm} \caption{
Average filled length $f$ and the wasted length $w$ for shrinking
particles as functions of the length $x$ of the adsorption
interval assumed initially empty. The assumed shrinking model is
particles randomly retracting their endpoints. The results are
obtained by iterating Eq. (9) with the assumed shrinking ratio of
$l=1.6$.} \label{3f}
\end{figure}

For particles whose final size is fixed it is natural to study the
mean number of particles $f(x)$ adsorbed in a line of length $x$. A
convenient equation for $f(x)$ valid for $x>1$ is obtained by
substituting $w=x-f(x)$ in Eq. (\ref{MastWst3}) and making the
replacement $x\rightarrow x+l$: \be f(x+l)=1+\frac{2}{x}\int_l^{max
\{x,l\}} f(y) +\frac{1}{x}\int_{max\{x,l\}}^{x+l-1} f(y) dy.
\label{eq-2ash}
\ee
In deriving Eq. (\ref{eq-2ash}) use has been made of the initial condition
\be f(x)=\left \{
\begin{matrix} 0,&\hfill&0\le x \le l \cr 1,&\hfill&l\le x \le
l+1 \cr\end{matrix} \right.  \label{bouCof}
\ee
Evaluation of $f(x)$ at small $x>1+l$ is readily done by a repeated
iteration procedure, going from the small to progressively larger length sizes. Results
of the numerical recursion are shown in Fig. 3. Note that already at
$x \approx 7$ the variations of $f(x)$ and $w(x)$ are very close to
linear.

An exact solution of Eq. (\ref{eq-2ash}), which will be used to
calculate variances, can be obtained using Laplace transformation.
Multiplying Eq. (\ref{eq-2ash}) by $x$, taking the Laplace
transform and using the initial conditions specified by Eq.
(\ref{bouCof}), we obtain an equation of the form \be -\frac
{d}{dp}\left(e^{pl}F(p)\right)=\frac{1}{p^2}+\frac{1}{p}F(p)+\frac{1}{p}e^{p(l-1)}F(p)
\label{suppl-1} \ee  for the  Laplace transform of $f(x)$, \be
F(p)=\int_0^\infty e^{-px} f(x)dx. \label{laplace} \ee Rearranging
terms and multiplying  by $e^{-pl}$, Eq. (\ref{suppl-1}) can be
rewritten in the form  \be
F'(p)+\left[l+\frac{1}{p}\left(e^{-pl}+e^{-p}\right)\right]F(p)=
-\frac{\exp(-pl)}{ p^2}. \label{eq-ff} \ee The solution of Eq.
(\ref{eq-ff}) satisfying the boundary condition at $p \rightarrow
\infty$, \be F(p)|_{p \rightarrow \infty}=\frac{1}{p}e^{-p~l},
\label{bound} \ee which follows from the known variation of $f(x)$
at small $x$, can be obtained in a straightforward manner: \be
F(p)=-\frac{ \exp(-p~l)}{p^2\beta_r(p)}\int_p^\infty \beta_r(u)du,
\label{eq-ff1} \ee where \be \beta_r(p)= \exp\left[- \int_0^{p}
\left(\frac{2- \exp(-v)-exp(-vl)}{v}\right)dv\right]. \label{beta}
\ee

In the analysis of the solution we follow the approach of Ref.
\cite{coff}, based on the use of Karamata's Tauberian theorem, see
e.g. \cite{Variat}, p. 37, for the asymptotic growth rate of
steadily growing functions. According to that theorem, in order to
obtain the asymptotic behavior of the filled length, $f(x)$ (or
the gap, $w(x)$, or of the variances of these functions) it is
sufficient to have a Laurent power-series expansion at small $p$
of the Laplace transforms of these functions (possibly cut at
small $x$ by a Heaviside step-function factor). Further
mathematical details of this type of analysis can be found in
\cite{coff}.

Function $F(p)$ is analytic at all $p\ne 0$ and has a second-order
pole at $p=0$ with the following expansion as $p \rightarrow 0$: \be
F(p)=\frac{\alpha_{f,0} }{p^2}+\frac{\alpha_{f,0} -1}{p}+O(p),
\label{asympF} \ee where \be \alpha_{f,0}=\int_0^\infty
\beta_r(p)dp. \label{filRen1} \ee To calculate $f(x)$ at large $x$,
we take the inverse Laplace transformation of the asymptotic
expansion (\ref{asympF}).  This gives \be
f(x)=\alpha_{f,0}x+\alpha_{f,0}-1, \label{solu-f} \ee with an
exponentially small error term. Whence we have
 \be w(x)=\alpha_{w,0}(x+1),  \ \
\alpha_{w,0}=1-\alpha_{f,0}. \label{solu-w} \ee

In the limit $l=1$, equation (\ref{filRen1}) gives the so-called
jamming filling factor $R$ for the standard RSA,
$\alpha_{f,0}(l=1) \equiv R=0.74759\cdots$ (also called the Renyi
constant \cite{renyi}). The filling factor (saturation coverage),
calculated with Eq. (\ref{filRen1}) as a function of the shrinkage
ratio  $l$, is shown in Fig. 4 (curve 2). As expected, the
shrinking causes a decrease of the filling factor with $l$.  Owing
to the fact that the correction terms are exponentially small, our
asymptotic solutions (\ref{solu-f}) and (\ref{solu-w}) are
extremely close (within less than $2 \cdot 10^{-4}$ for $x \ge 7$)
to the exact solution obtained by direct recursion (Fig. 3). This
accounts for the linearity of $f(x)$ and $w(x)$ at high $x$
evident in Fig. 3.

For particles shrinking to their centers, we use Eq. (\ref{MastW2})
and obtain \be f(x)=1+\frac{2}{x-l}\int_0^{x-l}
f\left(y+\frac{l-1}{2}\right)dy.
\label{eq-2ac}
\ee
Equation (\ref{eq-2ac}) can also be written in the form
\be
f\left(x+l+\frac{l+1}{2}\right)=1+\frac{2}{x+(l+1)/2}\int_l^{x+l}
f\left(y\right)dy.
\label{eq-2b}
\ee
Equation (\ref{eq-2b}) exhibits a recursion period $(l+1)/2$ which
is to be compared to a recursion period of unity for the standard RSA.
The initial condition for Eq. (\ref{eq-2b}) is given by
Eq. (\ref{bouCof}) and the resultant solution is similar in form to
Eq. (\ref{eq-ff1}) with the following replacement:
$\beta_r(p)\rightarrow \beta_c(p)$, where \be
\beta_c(p)= \exp\left[-2 \int_0^{p(l+1)/2} \left(\frac{1- \exp(-v)}{v}\right)dv\right].
\label{beta_c}
 \ee
\begin{figure} 
\epsfig{figure=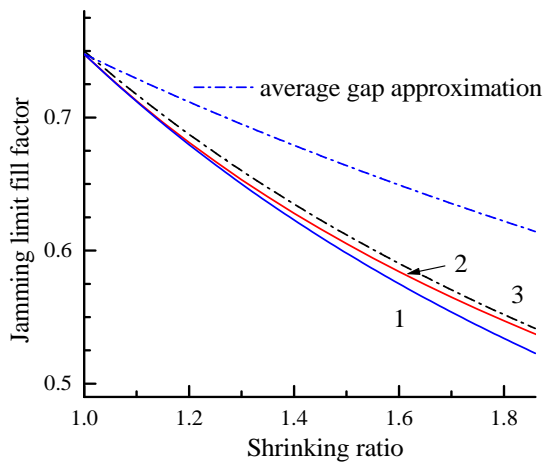,width=7.3cm,height=6.3cm} \caption{
Filling factor $\alpha_{f,0}(l)$ as function of the shrinking
ratio $l$ for different shrinking models. Curve (1) describes
particles shrinking to their centers. For the model of particles
randomly retracting their endpoints curve (2) shows the exact
solution and curve (3) the linear approximation. The upper curve
corresponds to the average-gap approximation.} \label{4f}
\end{figure}
From Eqs. (\ref{beta_c}) and (\ref{filRen1}) it follows that \be
\alpha_{f,0}=\frac{2}{1+l}R \label{FilN} . \ee Equation
(\ref{FilN})  represents the exact solution for this model in a
simple analytical form.  The fill factor calculated with Eq.
(\ref{FilN}) is shown in Fig. 4 as curve 1.  As can be expected,
the shrinking effect is smaller for particles shrinking by
randomly retracting their endpoints than for particles shrinking
to their centers.  This is due to the reduced contribution of edge
regions.

As noted above, the model of particle shrinking to the center
accounts for a strong repulsion between the particles acting as
hard disks. The minimal distance between the particle centers (an
effective particle size) equals to $(l+1)/2$. Therefore the number
of a unit size particles adsorbed in a line with fixed length and
the resulting jamming limit fill factor are reduced by $2/(l+1)$.

An alternative approximate way to obtain an asymptotic solution of
Eqs. (\ref{eq-2ash}), (\ref{eq-2b}) is to seek the solution in the
linear form $f(x)=ax+b$ in the region $x>nl+1$, for some $n$, while
using a function obtained by direct iterations for $x<nl+1$. This
recipe also gives a very good result starting from $n=1$. Thus, say,
for particles shrinking to centers we obtain
$\alpha_{f,0}^{(1)}=2/(1+l) R^{(1)}$, with $R^{(1)}=0.75$, quite
close to the exact Renyi constant. A less accurate approach is to
merely match the linear asymptotic to the recursion result for
$x<nl+1$.  In this case, one would need $n=2$ to obtain a similar
precision.

In calculations of the quantum yield and the Fano factor for the PEB
problem, the complexity of equations often inspires even more
radical approximations, based on estimates of the average losses,
see e.g. \cite{Spieler}. In terms of the RSA problem, this is
equivalent to assuming the filling factor $\alpha_{f,0}$ in the form
$\alpha_{f,0}=R(R+\alpha_{w,0})^{-1}$ with the wasted length growing
linearly with $l$, i.e. taking $\alpha_{w,0}=(1-R) \cdot l$.
Comparing the exact and the approximate solutions of Eq.
(\ref{eq-2ac}) (see Fig. 4), we see that this approach severely
underestimates the effect of shrinking.

{\bf Expanding particles.} We consider now the case when the initial
particle size is $a<1$. If the interval at the adsorbing line is
$a<x<1$, the particle fills it entirely (obviously, without
fluctuations). If $x>1$ the particle expands to the length of unity.
Evidently, for $x < a$ the whole length is wasted, $w(x)=x$. For $a
<x<1$ no waste occurs, while for $1<x<1+a$ the waste equals $w=x-1$
and does not fluctuate. For $x>a+1$ the probability of two particles
being adsorbed in this space steadily grows with $x$ and so does the
average waste. Thus, the initial conditions to Eq. (\ref{MastW3})
are of the form
\be w(x)=\left \{ \begin{matrix} x,&\hfill&0\le x
\le a \cr 0,&\hfill&a\le x \le 1 \cr x-1,&\hfill&1\le x \le 1+a
\cr\end{matrix} \right.  \label{bouCow} \ee

Fig. 5 shows the functions $w(x)$ and $f(x)$, obtained by direct
iterations of Eq. (\ref{MastW3}). Note a step-like feature in $f(x)$
and $w(x)$ at $x=1+a$ (which is replicated at $x=n+a$ with ever
smaller amplitude) and a decrease of the $w(x)$ in the region $x\ge
1+a$. These are ``recursive replicas'' of the singularity in OPDF
discussed above and the gap in $w(x)$ at $a<x<1$ .

\begin{figure} 
\epsfig{figure=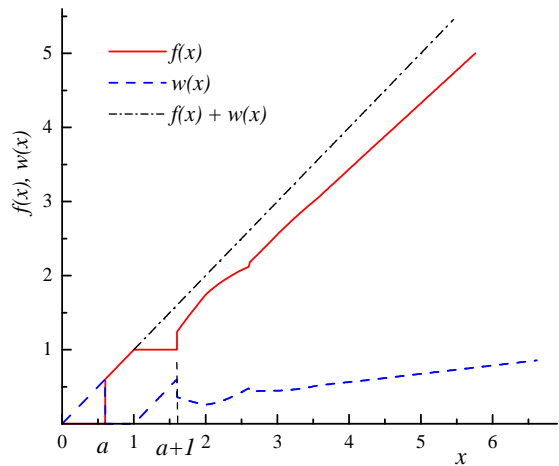,width=7.4cm,height=6.3cm} \caption{
Average filled length $f$ and the wasted length $w$ for expanding
particles as functions of the length $x$ of the adsorption
interval assumed initially empty. Results are evaluated by
recursion with Eq. (8) for an assumed expansion ratio of
$l=1/0.6$.} \label{5f}
\end{figure}

To calculate the variance we shall need an exact solution for
$w(x)$. It can be obtained by taking Laplace transformation of Eq.
(\ref{MastW3}).  First, it is convenient to  multiply Eq.
(\ref{MastU3}) by $x-a$ and make the substitution $x \rightarrow
x+1$. Taking Laplace transformation, we find that the function \be
F_w (p)=\int_1^\infty e^{-px} w(x)dx, \label{laplaceW} \ee (which
is the Laplace transform of $w(x)$ cut at small $x<1$ by
multiplying with a step function) satisfies the following equation
\ba \left [ - \frac
{d}{dp}+1-a\right]\left(e^{pl}F_w(p)\right)=\frac{2}{p}F_w(p)
\hspace{1.2cm} \nonumber \\ \hspace{1.0cm}
+(1-a)F_w(p)+\left(\frac{2}{p}+1-a\right)J(p), \label{suppl-n} \ea
where \be J(p)=\int_0^1w(x)e^{-px}. \label{suppl-T} \ee Using Eq.
(\ref{bouCow}) to calculate $J(p)$, we can re-write  Eq.
(\ref{suppl-n}) in the form  \be F'_w (p)+\left(a+\frac{2
e^{-p}}{p}b(ap)\right)F_w(p) =-\frac{\exp(-p)} {p^2}G_w(p),
\label{eq-W} \ee where \be G_w(p)=\frac{2}{p}b(ap)J_1(ap),
\label{g_w} \ee and \be b(ap)=1+\frac{1}{2}(1-a)p, \hspace{0.5cm}
\ J_1(ap)=\int_0^{ap}te^{-t}dt. \label{fact1}  \ee Solution of Eq.
(\ref{eq-W}) is of the form \be F_w (p)=-\frac{ \exp(-p)}{p^2
\tilde \beta(p)} \int_p^\infty \tilde \beta(u)G_w(u)du,
\label{eq-ffW} \ee where \be \tilde \beta(p)= \exp\left[-2
\int_0^{p} \left(\frac{1- e^{-v}}{ v}\right)b(av) dv\right].
\label{BetaExp} \ee Function $F_w (p)$ is analytic at all $p\ne 0$
and has a second-order pole at $p=0$. As $p \rightarrow 0$, the
following asymptotic expansion holds: \be F_w
(p)=\frac{\alpha_{w,0} }{p^2}+\frac{\alpha_{w,0}} {p}+O(p),
\label{asympPhi} \ee where now \be \alpha_{w,0}=\int_0^\infty
\tilde \beta(p)G_w(p)dp. \label{filWast} \ee Applying the inverse
Laplace transformation to Eq. (\ref{asympPhi}) we bring  $w(x)$ at
large $x$ into the form $w(x)=\alpha_{w,0}(x+1)$ with an
exponentially small error term. The filling factor
$\alpha_{f,0}=1-\alpha_{w,0}$ in the jamming limit is then given
by \be \alpha_{f,0}=2\int_0^\infty \tilde \beta(p) b(ap)\left(
\frac{e^{-ap}-e^{-p}}{p}+ae^{-ap}\right)dp \label{filExpa} \ee For
$a=1$, equation (\ref{filExpa}) again reduces to the Renyi
constant $R$.
\begin{figure} 
\epsfig{figure=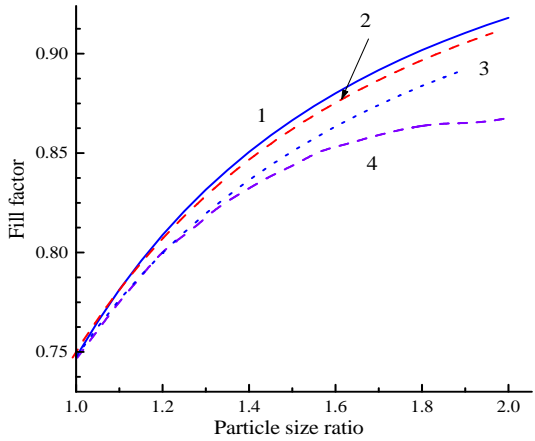,width=7.1cm,height=5.9cm} \caption{
Filling factor $\alpha_{f,0}$ as a function of the expansion ratio
$l=1/a$; curve (1) shows the exact result evaluated with Eq. (34),
curve (2) the approximate solution of Eq. (8), obtained by
linear-to-iterative matching, curve (3) describes the approximate
solution neglecting edge effects. Curve (4) corresponds to the
restricted expansion model of Ref. \cite{viot} (see also
Appendix).}\label{6f}
\end{figure}

The filling factor $\alpha_{f,0}$ calculated with Eq.
(\ref{filExpa}) is plotted in Fig. 6 as function of the expansion
ratio $l=1/a$. We see that expansion causes an increase of the
filling factor with $l$. Also shown are the results of Boyer et al.
\cite{viot} for the RSA problem with restricted particle expansion
(their model allows expansion only when the required gap is fully
available, see Appendix). As can be expected, the restricted expansion
gives a smaller increase of the fill factor.

Figure 6 also shows (curve 3) the approximate results obtained by
neglecting edge effects; in this approximation the expansion effect
is reduced. Much better results are obtained in the approximation
(shown by curve 2) obtained by using as a solution of Eq.
(\ref{MastW3}) a linear dependence at $x>n+a$ recursively combined
with the exact solution for $x<n+a$. Good agreement with the exact
result is obtained already for $n=1$.

\section{Variance and Fano Factor}

{\bf Shrinking particles.} We shall consider in detail the model of
particles shrinking by randomly retracting their endpoints, and then
briefly discuss the case of shrinking to the center.

Equation (\ref{MastU1}) can be readily transformed into an equation
for the expected value of the occupied length squared $v={\bf E}
f^2$. Master equation for $v(x)$ is of the form \ba v(x)=1+ \frac{2}{x-l}\int_l^{x-l} [v(y)+2f(x)]dy \nonumber\\
+\frac{2 }{x-l}\int_{x-l}^{x-1} [v(y)+2f(x)]dy \nonumber
\\ +\frac{2}{x-l}\int_{l}^{x-l}f(y) f(y-x-1)dy.
\label{eq-var1f}
\ea
In deriving Eq. (\ref{eq-var1f}) we used the initial conditions
(\ref{bouCof}). The Laplace transform $M(p)=\hat L (v(x))$ satisfies
\be M'(p)+\left[l+\frac{1}{p}\left(e^{-pl}+e^{-p}\right)\right]M(p)=-
\frac{ \exp(-pl)}{p^2}R_f(p), \label{eqM}
\ee where
\be
R_f(p)=1+2pF(p)\left(e^{p(l-1)}+1\right)+2p^2F^2(p)e^{p(l-1)}
\label{Rf}
\ee
with $F(p)$ defined by Eq. (\ref{eq-ff1}). The solution of Eq.
(\ref{eqM}) can be written in a form similar to Eq. (\ref{eq-ff1}),
namely
\be
M(p)=-\frac{\exp(-p)}{p^2\beta_r(p)}\int_p^\infty
\beta_r(u)R_f(u)du. \label{eq-ffM} \ee The main feature of this
solution is the divergence of the integral in the right-hand side as
$p \rightarrow 0$, owing to the square-law dependence of the
variance on $x$ at large $x$. This singularity should be treated
with care.

To do this we note that $F(p) \propto \alpha_{f,0}p^{-2}$ at small
$p$ and hence the difference $2p^2F^2(p)-2\alpha_{f,0}^2p^{-2}$ is
regular at $p\rightarrow 0$. Therefore, it is convenient to define
an entire function
$\kappa(p)=\beta(p)R_f(p)-2\alpha_{f,0}^2p^{-2}$. In terms of this
function the solution $M(p)$ can be expressed as follows: \be
M(p)=-\frac{ \exp(-p)}{p^2\beta(p)}\left[\frac{2\alpha^2_{f,0}}{p}
+k_{f,0}-\int_0^p \kappa(u)du\right]. \label{eq-ffren} \ee The
asymptotic expansion of $M(p)$ near its pole of the third order is
of the form \be M(p)=\frac{2\alpha_{f,0}^2}{p^3}+\frac{k_{f,0}+2
\alpha_{f,0}^2}{p^2} +\frac{\alpha_{f,0}^2}{p}. \label{series-M}
\ee Applying the inverse Laplace transformation, we find the
asymptotic form of $v(x)$: \be v(x) = \alpha_{f,0}^2 x^2+
(k_{f,0}+2 \alpha_{f,0}^2) x+k_{f,0}+ \alpha_{f,0}^2+1 ,
\label{FFF} \ee with an exponentially small error term.
Subtracting $f^2(x)$, we find $v(x)-f^2(x)=\mu_r(x+1)$ where
$\mu_r=k_{f,0}+2 \alpha_{f,0}$ is the specific variance of the
fill factor (at $x\rightarrow \infty$), \ba \mu_r=3\alpha_{f,0}+
2\int_0^\infty \beta_r(u)pF(u)(e^{p(l-1)}+1)du \nonumber\\
+2\int_0^\infty \left[\beta_r(u)u^2F^2(u)e^{p(l-1)}
-\frac{\alpha_{f,0}^2}{u^2}\right]du. \ea Integrating by parts the
second term and rearranging the result, we rewrite $\mu_r$ in the
form \ba \mu_r=2\int_0^\infty \frac{\alpha_f(u)}{u}\left(e^{-u}
+e^{-ul}-2e^{-u(l+1)}\right)du\nonumber \\-2\int_0^\infty
\frac{\alpha^2_f(u)}{\beta_r(u)u^2}
e^{-u(l+1)}\left((l+1)u+e^{-u}+e^{-ul}-2\right)du \nonumber\\
-\alpha_{f,0},\hspace{2cm} \label{var_k} \ea where
\be
\alpha_f(u)=\alpha_{f,0}-\int_0^u \beta_r(y)dy \label{Alp_u}.
\ee

In the limit $l=1$, Eq. (\ref{var_k}) reduces to the standard RSA
result first obtained for a lattice RSA model by Mackenzie
\cite{Mack}. The numerical value of the Mackenzie constant $\mu_0 =
0.0381564..$ corresponds to the Fano factor $\Phi =0.0510387..$ (see
\cite{coff} for detailed estimates of the variance). Expression
(\ref{var_k}) for shrinking particles has the same structure as the
corresponding formula for the standard RSA model (fixed-size CPP).

Specific variance $\mu_r$ of the filling factor obtained from Eq.
(\ref{var_k}) is plotted in Fig. 7 against the excess parking length
factor $l$. We also plot the ratio of the filling factor variance to
its mean value $\mu_r/\alpha_{f,0}$, known in the high-energy
detector physics \cite{Fano,rusb} as the Fano factor $\Phi$. We see
that while the fill factor variance decreases with $l$ (as does the
mean fill factor itself, cf. Fig. 3), their ratio $\Phi$ increases
with $l$ due to the steeper decrease of the mean fill factor.

\begin{figure} 
\epsfig{figure=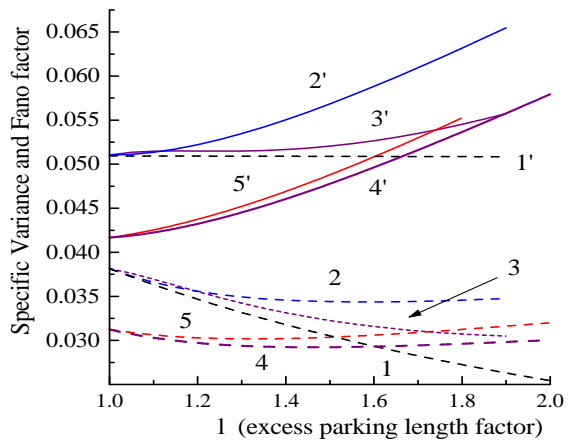,width=7.4cm,height=5.9cm} \caption{
Specific variance  and the Fano factor (shown by curves with
primes) as functions of the shrinking ratio $l$. Shown are results
of exact calculations: (1,1$^{\prime}$) - scaled results for
particles shrinking to center, (2,2$^{\prime}$)- results for
particles randomly shrinking to the ends [based on Eq.
(\ref{var_k})],
 and (3,3$^{\prime}$)-  results of  solution neglecting edge effects; also show results
 of approximation for $f(x)$ and $u(x)-f(x)^2$ at $x>l+1$ allowing edge effect (4,4$^{\prime}$)
 and without it (5,5$^{\prime}$).}\label{7f}
\end{figure}

Similar analysis of the model in which particles shrink to the
center shows that for this case the variance asymptotic is given
by a constant $\mu_c=2/(l+1)\mu_0$, so that it decreases as fast
as fill factor. As a result, the Fano factor is not modified in
this model. This is in line with the fact that both the number of
adsorbed particles and their distribution is identical to that for
a standard RSA with the modified particle size.

Figure 7 also shows the variance and the Fano factor obtained by an
approximate solution of Eq. (\ref{MastU2}) based on linear
approximations for both the mean waste length $w(x)=ax+b$ and the
variance $u(x)-w(x)^2=cx+d$ for $x>1+l$ recursively combined with
exact expressions in the region $x<1+l$. We see that the difference
between the approximate and exact solutions is quite substantial.
Moreover, the approximate solution exaggerates the growth of the
Fano factor with $l$.

{\bf Expanding particles.} In this case it is convenient to solve
equation (\ref{MastU3}) for the gap (waste) variance, using Eq.
(\ref{bouCow}) for the boundary conditions.  We define a function
\be N(p)=\int_1^\infty e^{-px}u(x)dx  \label{four-u} \ee  that
satisfies an equation of the same type as Eq. (\ref{eq-W}), in
which one should replace $G_w(p) \rightarrow R_w(p)$, where \be
R_w(p)= \frac{2J_2(ap)}{p^2}b(ap)  + 2p^2 \left(F_w(p) +
\frac{1}{p^2} J_1(ap)\right)^2, \label{R_w} \ee with \be
J_2(ap)=\int_0^{ap}t^2e^{-t}dt. \label{Ineg-s} \ee Similarly to
the standard RSA case,  $N(p)$ has a third-order pole, i.e. at
$p\rightarrow 0$ \be
N(p)=\frac{\alpha_{w,0}^2}{p^{3}}+\frac{(k_{w,e}+\alpha_{w,0})^2}{p^{2}}
+\frac{k_{w,e}+\alpha_{w,0}}{p}. \label{N-ser} \ee Separating out
the regular part of $N(p)$ and rearranging the terms, one obtains
the constant $k_{w,e}$ in the form $k_{w,e}=K_1+K_2+K_3$, where
\be K_1=2\int_0^\infty \frac{\tilde \beta(u)}{u^2}\left(
b(au)J_1^2(au)+J_2(au) \right) du, \label{k-1} \ee \be
K_2=4\int_0^\infty \frac{\alpha_w(u)}{u^2} e^{-2u}
\left(1-2e^{-u}b(au)\right)J_1(au)du, \label{k-2} \ee and \be K_3=
-4\int_0^\infty \frac{\alpha_w(u)^2}{\tilde \beta(u)u^2}
e^{-u}\left((e^{-u}-1)b(au)+u\right) du. \label{k-3} \ee Here \be
\alpha_w(u)=\alpha_{w,0}-2\int_0^u\frac{\tilde \beta(p)}{p}
b(ap)J_1(ap)dp. \label{alpha_w} \ee
\begin{figure} 
\epsfig{figure=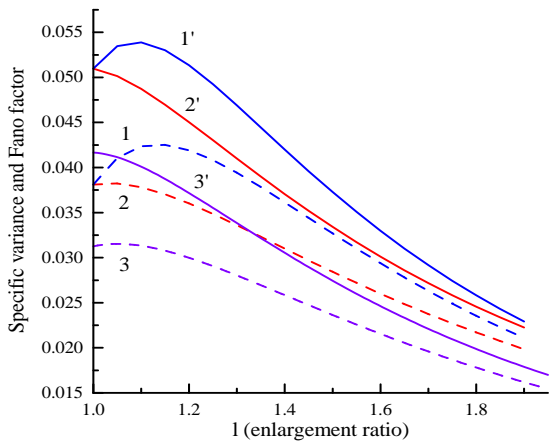,width=7.4cm,height=5.9cm} \caption{
Specific variance (dashed curves) and the Fano factor (solid
curves with primed labels) as functions of the expansion ratio
$l=1/a$. Exact calculations with Eqs. (48-50) are shown by curves
(1,1$^\prime$), curves (2,2$^\prime$) correspond to edge effect
neglected, and curves (3,3$^\prime$) represent the linear
approximation of $w(x)$ and $u(x)-w(x)^2$ at $x>l+1$.}\label{8f}
\end{figure}

Variance of the gap distribution is obtained by the inverse
Laplace transformation of $N(p)$. At large $x$ one has
$v(x)-w^2(x)=k_{w,e}(x+1)$. The constant $k_{w,e}$ describes, with
exponentially small errors, the linear dependencies of both the
filled length and the gap variance.

With the substitution $\alpha_w(u)=(u+1)\tilde \beta(u)-\alpha_f(u)$
and integration by parts, formulae (\ref{k-1}-\ref{k-3}) can be
rearranged after some lengthy algebra into a form similar in
structure to Eq. (\ref{var_k}). However, additional terms at $a\ne
1$ make it rather unwieldy. Both forms are equivalent for numerical
integration.

Variance of the gap distribution calculated with Eqs.
(\ref{k-1}-\ref{k-3}) equals the filled length variance. Figure 8
displays specific variance along with the Fano factor as functions of
$l=1/a$.  Note the increase of both the variance and the Fano factor
at small $l$. This increase is apparently due to the singular
contribution to OPDF from edge effect. For comparison, we show the
$v(x)$ and $\Phi$ calculated neglecting the edge effect (this is
accomplished by taking $b(ap)=1$):  no increase is seen in this
approximation. Both the variance and the Fano factor vanish with
increasing $l$ owing to the fact that variably growing particles
provide tighter filling.

Figure 8 also shows the variance and the Fano factor obtained by an
approximate solution of Eq. (\ref{MastU3}) based on the linear
approximations for both the mean waste length $w(x)=ax+b$ and the
variance $u(x)-w^2(x)=cx+d$ for $x>1+l$ combined with exact
expressions for $x<1+l$. In contrast to the case of shrinking
particles, the difference between the approximate and the exact
solutions becomes smaller at larger expansion ratios, since both
solutions predict a rapid decrease of the Fano factor with $l$. Note
that in this case the overall variation is well described by the
approximate solution.

\section{conclusions}
We have considered a generalized 1-dimensional random sequential
adsorption problem, where particles shrink or expand upon
adsorption. Using a recursive approach, we obtained exact analytical
expressions for the filling factor and its variance.

In the model where particles shrink to their centers, we find that
both the filling factor and its variance scale as the recursion
period, so that the Fano factor remains unchanged. But in another
model of shrinking, where particles shrink by randomly retracting
their endpoints, the Fano factor increases with larger shrinkage
ratio, due to the weaker decrease of the variance compared to that
of the filling factor. In the case of adsorption of expanding
particles (within the allowed intervals), we find that the variance
first increases with the expansion ratio $l$ and then decreases,
showing a maximum at relatively small values of $l$. These
nontrivial results are due to edge effects, resulting in a modified
one-particle gap distribution function, $\rho(y)$, near the edges of
the adsorbing interval.

The developed approach can be applied to other models, so long as
they correspond to a  piecewise constant $\rho(y)$. The results are
shown to have exponentially small corrections for finite adsorption
length $x$; for $x > 7\hspace{1 mm}l $ they become practically
exact.

The analytic results have been compared to approximate solutions
based on a method frequently used in semiconductor physics. The
approximate solution is reasonably accurate in estimating the fill
factor for either shrinking or expanding particles. However, the
same method is less accurate in estimates of the fill factor
variance. In the model where particles shrink by retracting their
endpoints the approximate method overestimates the effects of
shrinkage on both the variance and the Fano factor. We also assessed
another common approximate method, based on the mean final energy
distribution approach, and found that it gives only qualitative
trends.

Our quantitative results have important applications to the problem
of energy branching in high-energy particle propagation through a
semiconductor crystal, where the model of shrinking particles (where
particles shrink symmetrically about their centers) naturally
accounts for the fact that the impact ionization threshold is larger
than the energy gap for electron-hole pair generation. The
alternative model of shrinking particles, where shrinking occurs by
retracting one of the particle endpoints, further accounts for the
decreasing density of states at low particle energies.

{\bf Acknowledgement.} This work was supported by the New York State
Office of Science, Technology and Academic Research (NYSTAR) through
the Center for Advanced Sensor Technology (Sensor CAT) at Stony
Brook.

\appendix
\section{Restricted expansion model, comparison with rate equation results}
\label{Appendix2} In the model of RSA with expanding particles
studied in \cite{viot}, the authors considered the case of
restricted growth, when the deposited particle begins to grow only
if the interval in which it is adsorbed is larger than its final
size. For this model, Eq. (\ref{MastU3}) remains valid for intervals
$x>1$, but the initial conditions to Eq. (\ref{MastU3}) for $w(x)$
(at $x<1$) are of the form \be w(x)=\left \{ \begin{matrix}
x,&\hfill&0\le x \le a \cr x-a,&\hfill&a\le x \le 1 \cr
x-1,&\hfill&1\le x \le 1+a \cr\end{matrix} \right.  \label{bouCon2}
\ee To solve this model, we shall use the same technique as in
Section \ref{fillFx}, so as to compare the results of the kinetic
and the recursive approaches. To do this, we multiply  Eq.
(\ref{MastU3}) by $x-a$, make the substitution $x \rightarrow x+1$
and then take Laplace transformation. We obtain an equation similar
in form to Eq. (\ref{suppl-n}) where, however, $J(p)$ must be
calculated using the revised initial conditions of Eq.
(\ref{bouCon2}). This results in \be J(p)=\int_0^1w(x)e^{-px}
=\frac{1}{p^2}\left[J_1(ap)+e^{-ap}J_1\left((1-a)p\right)\right].
\label{suppl-T} \ee The solution of Eq. (\ref{suppl-n}) and hence
$\alpha_{w,0}$ are of the form similar to Eqs. (\ref{eq-ffW}) and
(\ref{filWast}) in which one must use $G_w(p)$ defined by Eq.
(\ref{fact1}) but with $J_1(ap)$ replaced by the sum of two terms in
the square brackets of Eq. (\ref{suppl-T}). Calculating the filling
factor as $\alpha_{f,0}=1-\alpha_{w,0}$, we find \be
\alpha_{f,0}=2\int_0^\infty \tilde \beta(p) b(ap)\left[ a
e^{-ap}+(1-a)e^{-p}\right] dp \label{filBoy}. \ee In Ref.
\cite{viot}, the contribution of expanded and non-expanded particles
was calculated separately for a finite time $t$, using the rate
equation approach. In order to make the comparison, one has to take
an appropriate linear combination (Eq. (15) and Eqs. (31), (32) of
Ref. \cite{viot}) in the limit of infinite time. This yields (in the
notation of \cite{viot}) the fill factor $\theta$ in the form: \be
\theta=\int_0^\infty
F(t)\left\{1+(\sigma-1)\left[1+(\sigma-1)t\right]e^{-(\sigma-1)t}\right\}dt.
\label{origBoy} \ee To compare Eqs. (\ref{filBoy}) and Eq.
(\ref{origBoy}), we multiply the right-hand side integrand in
(\ref{filBoy}) by unity in the form $1-e^p+e^p$. Then, the
contribution $I$, proportional to $1-e^p$, can be rewritten as \be
I=\int_0^\infty d \left(\tilde \beta(p)\right) p \left[ a
e^{(1-a)p}+(1-a)\right] \label{I-1} \ee and then integrated by parts
to give \be I=- \int_0^\infty  \tilde \beta(p) \left\{
a\left[1+(1-a)p\right] e^{(1-a)p}+1-a \right\}dp. \label{I-1} \ee
The remaining part $II$ is given by \be II=\int_0^\infty  \tilde
\beta(p) \left[ a e^{(1-a)p}+(1-a)\right]\left[2+(1-a)p\right] dp.
\label{I-2} \ee The sum $I+II$ gives $\alpha_{f,0}$ in the form \be
\alpha_{f,0}=\int_0^\infty \tilde \beta(p) \left[ a
e^{(1-a)p}+(1-a)(1+(1-a)p\right] dp. \label{filFinB} \ee Equation
(\ref{filFinB}), after the substitutions  $\tilde
\beta(p)=F(p)\exp[{-(1-a)p}]$, $a=1/\sigma$ and $p= \sigma t$,
transforms into an expression identical to the $\theta$ given by Eq.
(\ref{origBoy}) derived from \cite{viot}.

It should be noted, however, that while the fill factors are
identical, success of the kinetic approach is hard to extend to
variance calculations, where it appears to be much less effective
(see e.g. \cite{Talbot}).

\begin{thebibliography}{99}
\bibitem{gamma} The basic Eqs. (14), (15) of Ref. \cite{rusb} for the high energy particle
energy waste and its variance reduce to the RSA model equations
for the gap and its variance for the value of parameter
$\gamma=0$, which means that one of the 3 secondary particles has
a small kinetic energy.
\end{thebibliography}
\end{document}